\documentclass[final,5p,times,twocolumn]{elsarticle}

\usepackage{mathtools}
\usepackage{braket} 
\usepackage[dvipsnames]{xcolor}  
\usepackage{float}
\usepackage[T1]{fontenc}
\usepackage[utf8]{inputenc}
\usepackage{graphicx,color,rotating,pifont}
\usepackage{amsmath,amssymb}
\usepackage{mathrsfs}
\usepackage{mathtools}
\usepackage{subfigure}
\usepackage{siunitx}
\usepackage{bm}
\usepackage{ae}
\usepackage{dcolumn}
\usepackage{txfonts}
\usepackage{tensor}
\usepackage{braket}
\usepackage{booktabs}
\usepackage{xspace}
\usepackage{soul}
\usepackage{multirow}
\usepackage{tikz}
\usepackage{pgffor} 
\setstcolor{red}
\setul{}{.2ex} 
 
\usepackage[normalem]{ulem}

\usetikzlibrary{backgrounds}
\usetikzlibrary{shapes,matrix,trees}
\usetikzlibrary{arrows.meta}
\usetikzlibrary{positioning}			
\usetikzlibrary{calc,through}			
\usetikzlibrary{positioning,calc,shapes,decorations.pathreplacing,decorations.markings,decorations.pathmorphing}
\tikzset{
	vector/.style={decorate, decoration={snake,amplitude=2.5pt}, draw},
	provector/.style={decorate, decoration={snake,amplitude=2.5pt}, draw},
	antivector/.style={decorate, decoration={snake,amplitude=-2.5pt}, draw},
	fermion/.style={draw=black, postaction={decorate},
		decoration={markings,mark=at position .6 with {\arrow[draw=black]{>}}}},
           vL/.style={draw=ppurple, postaction={decorate},
		decoration={markings,mark=at position .6 with {\arrow[draw=ppurple]{>}}}},
	vLp/.style={draw=ppurple, postaction={decorate},
		decoration={markings,mark=at position .7 with {\arrow[draw=ppurple]{>}}}},
	NR/.style={draw=ggreen, postaction={decorate},
		decoration={markings,mark=at position .62 with {\arrow[draw=ggreen]{>}}}},	
	NRp/.style={draw=ggreen, postaction={decorate},
		decoration={markings,mark=at position .7 with {\arrow[draw=ggreen]{>}}}},	
	neutralino/.style={draw=black},
	fermionbar/.style={draw=black, postaction={decorate},
		decoration={markings,mark=at position .6 with {\arrow[draw=black]{<}}}},
	fermionnoarrow/.style={draw=black},
	gluon/.style={decorate, draw=black,
		decoration={coil,amplitude=4pt, segment length=5pt}},
	scalar/.style={dashed,draw=black, postaction={decorate},
		decoration={markings,mark=at position .55 with {\arrow[draw=black]{>}}}},
	scalarbar/.style={dashed,draw=black, postaction={decorate},
		decoration={markings,mark=at position .55 with {\arrow[draw=black]{<}}}},
	scalarnoarrow/.style={dashed,draw=black},
	electron/.style={draw=black, postaction={decorate},
		decoration={markings,mark=at position .55 with {\arrow[draw=black]{>}}}},
	bigvector/.style={decorate, decoration={snake,amplitude=4pt}, draw},
	photon/.style={decorate, draw=black,decoration={snake,amplitude=4pt, segment length=5pt} }
}

\usepackage[colorlinks,linkcolor=blue,anchorcolor=blue,citecolor=blue]{hyperref} 
\hypersetup{
   colorlinks=true, 
   linkcolor=blue,  
   citecolor=blue,  
   filecolor=blue,  
   urlcolor=blue    
} 

\DeclareMathSymbol{\NS}{\mathord}{AMSb}{"4E}

\DeclareSIUnit{\fm}{\femto\meter}

\newcommand{\nuc}[2]{\ensuremath{{}^{#2}\mathrm{#1}}}

\newcommand{\beq}{\begin{equation}}
\newcommand{\eeq}{\end{equation}}
\newcommand{\beqn}{\begin{eqnarray}}
\newcommand{\eeqn}{\end{eqnarray}}
\newcommand{\bsub}{\begin{subequations}}
\newcommand{\esub}{\end{subequations}}
\newcommand{\bpm}{\begin{pmatrix}}
\newcommand{\epm}{\end{pmatrix}}

\newcommand\identity{1\kern-0.25em\text{l}}

\newcommand{\onbb}{0\nu\beta\beta}
\def\nn{\nonumber}

\makeatletter

\newcommand{\Rmnum}[1]{\expandafter\@slowromancap\romannumeral #1@}
\makeatother

\begin{document}

\begin{frontmatter}

\title{Nuclear matrix elements of neutrinoless double-beta decay in covariant density functional theory with different mechanisms}

\author{Chen-rong Ding\fnref{1}}
\author{Gang Li\fnref{1}} 
 \author{Jiang-ming Yao\fnref{1}\corref{cor1}}
    \ead{yaojm8@sysu.edu.cn}
    \cortext[cor1]{Corresponding author}
\address[1]{School of Physics and Astronomy, Sun Yat-sen University, Zhuhai 519082, P.R. China}

\date{\today}

\begin{abstract} 
Nuclear matrix elements (NMEs) for neutrinoless double-beta ($0\nu\beta\beta$) decay in candidate nuclei play a crucial role in interpreting results from current experiments and in designing future ones. Accurate NME values serve as important nuclear inputs for constraining parameters in new physics, such as neutrino mass and the Wilson coefficients of lepton-number-violating (LNV) operators. In this study, we present a comprehensive calculation of NMEs for $0\nu\beta\beta$ decay in $^{76}$Ge, $^{82}$Se, $^{100}$Mo, $^{130}$Te, and $^{136}$Xe, using nuclear wave functions obtained from multi-reference covariant density functional theory (MR-CDFT). We employ three types of transition potentials at the leading order in chiral effective field theory. Our results, along with recent data, are utilized to constrain the coefficients of LNV operators. The results demonstrate that the combined NMEs based on the Feynman diagrams at the hadronic scale for the nonstandard mechanisms leads to the uncertainty by different nuclear models comparable to that for the standard mechanism. The use of NMEs from various nuclear models does not dramatically change the parameter space intervals for the coefficients, although MR-CDFT yields the most stringent constraint. Furthermore, our NMEs can also be used to perform a more comprehensive analysis with multiple isotopes.

\end{abstract}
  
\begin{keyword}

Neutrinoless double-beta decay\sep
Covariant density functional theory \sep
Nuclear matrix elements 

\end{keyword}

\end{frontmatter} 

\section{Introduction}

Neutrinoless double-beta ($0\nu\beta\beta$) decay is a hypothetical second-order weak-interaction process in which an even-even nucleus decays into its neighboring even-even nucleus with the emission of only two electrons~\cite{Furry:1939}. The observation of this process would provide direct evidence for the existence of lepton-number-violating (LNV) processes in nature and implies the existence of a Majorana mass term for the neutrino~\cite{Schechter:1982PRD,Haxton1984PPNP}, despite the existence of other interpretations~\cite{Burgess:1992dt,Burgess:1993xh,Huang:2021kam,Graf:2023dzf}. Therefore, the search for $0\nu\beta\beta$ decay in atomic nuclei has become a significant research frontier in particle and nuclear physics~\cite{Dolinski:2019,Agostini:2022RMP,Adams:2022White_Paper,Cirigliano:2022JPG,Cirigliano:2022Snowmass}. 

In addition to the standard mechanism of exchanging light Majorana neutrinos, there are several nonstandard $0\nu\beta\beta$-decay  mechanisms~\cite{Rodejohann:2011IJMPE}.
Many studies related to the nonstandard mechanisms~\cite{Doi:1985,Mohapatra:1986,Vergados:1987,Tomoda:1991,Hirsch:1995PRL,Hirsch:1996PLB,Simkovic:2010PRD,Deppisch:2012nb,Li:2021PRL,deVries:2022JHEP,Patra:2023PRD,Fukuyama:2023,Bolton:2021hje} are based on specific new physics models, including the R-parity violating supersymmetric model~\cite{Dreiner:1997uz,Allanach:2003eb,Barbier:2004ez}  and the left-right symmetric model~\cite{
Mohapatra:1980PRL,Mohapatra:1981PRD}. In both cases, the nonstandard mechanisms can be categorized into long-range  and short-range ones.  For instance, in the left-right symmetric model, apart from the standard mechanism, the $0\nu\beta\beta$ decay
receives nonstandard contributions from the exchange of either left-handed neutrinos or right-handed neutrinos, which depend on the momentum transfer rather than the light Majorana neutrino masses\footnote{In the literature, this is also called $\lambda$ and $\eta$ mechanisms for different chiralities of quarks~\cite{Doi:1982dn,Barry:2013xxa}.}, or the right-handed neutrino mass, respectively. The related nuclear matrix elements (NMEs) have been determined using nuclear wave functions from quasiparticle random approximation (QRPA)~\cite{Muto:1989ZPA,Faessler:1998JPG,Hyvarinen:2015,Stefanik:2015PRC,Simkovic:2017FP} and interacting shell models (SM)~\cite{Horoi:2016PRD,Menendez:2017fdf,Ahmed:2017PLB,Ahmed:2020PRD}. 

More generally, the long-range part~\cite{Pas:1999} and short-range  part~\cite{Pas:2001PLB} of the $0\nu\beta\beta$-decay rate could be parameterized in terms of different effective couplings involving standard and nonstandard currents allowed by Lorentz invariance. Within this framework, the NMEs of calculations with the wave functions from different nuclear models are employed to set limits for the coupling constants of arbitrary LNV operators, including those in some particle physics scenarios~\cite{Deppisch:2012JPG,Horoi:2017gmj,Deppisch:2020PRD,Kotila:2021longrange,Chen:2024}.

The effective field theory (EFT) provides a systematical and model-independent way to describe $0\nu\beta\beta$-decay rate from new physics scale to nuclear physics scale~\cite{Cirigliano:2022Snowmass}. A {\em master formula} of the half-life for the $0\nu\beta\beta$ decay was proposed in Refs.~\cite{Cirigliano:2017,Cirigliano:2018JHEP}, where all LNV operators up to dimension nine have been studied following the pioneering work~\cite{Prezeau:2003PRD}. It turns out that the contact LNV operators at quark level would induce a chirally enhanced contribution from pion exchange at the hadronic scale in chiral EFT. Most recently, this framework has been extended further by incorporating sterile neutrinos in Ref.~\cite{Dekens:2020JHep}. 
This framework has recently attracted significant attention in the nuclear physics community because it helps quantify the uncertainty of NMEs from transition operators~\cite{Cirigliano:2022JPG}. This was demonstrated in a recent {\em ab initio} study on the NMEs of transition operators, which rapidly converges with the order of chiral expansion~\cite{Belley:2024PRL}.

In this work, we exploit the EFT framework to study  the contributions to $0\nu\beta\beta$ decay from three mechanisms, namely, the standard, long-range, and short-range mechanisms. To this end, we arrange the non-relativistic reduced leading-order (LO) transition potentials in terms of a set of phenomenological LNV coefficients at the hadronic scale.  
The NMEs for the $\onbb$ decay in $^{76}$Ge, $^{82}$Se,  $^{100}$Mo, $^{130}$Te, and $^{136}$Xe are computed using nuclear wave functions from the multi-reference covariant density functional theory (MR-CDFT), which has been successfully employed in the studies of the NMEs of $\onbb$ decay based on the transition operators of the light~\cite{Song:2014,Yao:2015,Yao:2016PRC} and heavy Majorana neutrino exchange mechanisms~\cite{Song:2017,Ding:2023}.  To the best of our knowledge, the NMEs of candidate nuclei have not been computed using \emph{ab initio} methods based on nonstandard mechanisms. Therefore, it is important to compare the NMEs from different phenomenological models to examine the discrepancies among them in constraining the LNV operators. Our methodology, in particular, facilitates a more reasonable comparison of the NMEs for nonstandard mechanisms across different nuclear models and provides a convenient and systematic way of estimating their impact on the theoretical interpretation of the experimental limit on the $0\nu\beta\beta$ decay half-life.

\section{Theoretical framework of $0\nu\beta\beta$ decay} 
\label{sec:formalism}
 
The inverse half-life $(T^{0\nu}_{1/2})^{-1}$ of the $0^{+}\to 0^{+}$ $0\nu\beta\beta$ decay can be factorized as below~\cite{Doi:1985,Bilenky:2015,Cirigliano:2018JHEP,Yao:2022PPNP},
\begin{equation}
\label{eq:half-life-1}
\begin{aligned}
    \left( T^{0\nu}_{1/2}\right)^{-1}=&\frac{1}{8\ln2}\frac{1}{(2\pi)^5}\int \frac{d^3k_1}{2E_1}\frac{d^3k_2}{2E_2}|\mathcal{A}|^2\\
    &\times F(Z,E_1)F(Z,E_2)\delta(E_1+E_2+E_f-E_i),
\end{aligned}
\end{equation}
where the Fermi functions $F(Z,E_{1(2)})$ describe the distortion of electron wave functions from plane wave functions due to the presence of Coulomb interaction generated by the protons inside the nucleus~\cite{Cirigliano:2018JHEP}, where $Z$ is the atomic number of daughter nucleus.  $E_i, E_f$, and $E_{1(2)}$  are the energies of initial and final nuclear states, and electrons, respectively. $\mathcal{A}$ is the total amplitude for $0\nu\beta\beta$ decay. Similar to Ref.~\cite{Cirigliano:2018JHEP}, the amplitude is decomposed into different leptonic structures multiplied by the corresponding sub-amplitudes,
\begin{equation}
\begin{aligned}
\label{eq:totamplitude}
    \mathcal{A}=\frac{g_A^2 G_F^2 m_e}{\pi R_A}&\left[{\mathcal{A}_{L}\bar{u}\left(k_{1}\right) P_{R} C \bar{u}^{T}\left(k_{2}\right)+\mathcal{A}_{R}\bar{u}\left(k_{1}\right) P_{L} C \bar{u}^{T}\left(k_{2}\right)}\right.\\
    &\left.{+\mathcal{A}_{E}\bar{u}\left(k_{1}\right) \gamma_0 C \bar{u}^{T}\left(k_{2}\right)\frac{E_1-E_2}{m_e}}\right.\\
    &\left.{+\mathcal{A}_{m_e}\bar{u}\left(k_{1}\right)  C \bar{u}^{T}\left(k_{2}\right)+\mathcal{A}_{M}\bar{u}\left(k_{1}\right) \gamma_0\gamma_5 C \bar{u}^{T}\left(k_{2}\right)}\right].
\end{aligned}
\end{equation}
Here, $g_A$ is the $\pi N$ axial-vector coupling constant, $G_F$ the Fermi coupling constant, $m_e$ the electron mass, $P_{R(L)}$ are right- and left-handed projection operators, and $R_A=1.2A^{1/3}$ fm, where $A$ is the mass number of atomic nucleus.  $u(k_1)$ and $ u(k_2)$ denote the wave functions of two outgoing electrons with the momenta $k_1$ and $k_2$, respectively. Substituting (\ref{eq:totamplitude}) into (\ref{eq:half-life-1}), one finds 
\begin{align}
\label{eq:half-life-2}
\left(T_{1 / 2}^{0 \nu}\right)^{-1}= & g_A^4\left\{G_{01}\left(\left|\mathcal{A}_L\right|^2+\left|\mathcal{A}_R\right|^2\right)-2\left(G_{01}-G_{04}\right) \operatorname{Re} \mathcal{A}_L^* \mathcal{A}_R\right.\nonumber\\
&+4 G_{02}\left|\mathcal{A}_E\right|^2 +2 G_{04}\left[\left|\mathcal{A}_{m_e}\right|^2+\operatorname{Re}\left(\mathcal{A}_{m_e}^*\left(\mathcal{A}_L+\mathcal{A}_R\right)\right)\right]\nonumber\\
&-2 G_{03} \operatorname{Re}\left[\left(\mathcal{A}_L+\mathcal{A}_R\right) \mathcal{A}_E^*+2 \mathcal{A}_{m_e} \mathcal{A}_E^*\right]\nonumber\\
&\left.+G_{09}\left|\mathcal{A}_M\right|^2+G_{06} \operatorname{Re}\left[\left(\mathcal{A}_L-\mathcal{A}_R\right) \mathcal{A}_M^*\right]\right\}\;,
\end{align}
where the expressions of phase space factors $G_{0i}$ $(i=1-4,6,9)$ can be found in Refs.~\cite{Cirigliano:2018JHEP,Scholer:2023}.  The sub-amplitudes $\mathcal{A}_i$ with $i\in\{L, R, E, m_e, M\}$ distinguishing different leptonic structures  are defined as follows,
\beqn
\label{eq:subamplitude}
    \mathcal{A}_i
    &=& 4 \pi R_A \sum_a \bra{\Psi_F(0^{+})}\nonumber\\
    &&  \sum_{m,n}\int\frac{d^3\mathbf{q}}{(2\pi)^3}e^{i\mathbf{q}\cdot\mathbf{r}_{mn}}V^a_i(\mathbf{q}^2)\left(\tau^{+}_m \tau^{+}_n\right)\ket{\Psi_I(0^{+})},
\eeqn
where $\tau^{+}$ is the isospin raising operator, $\mathbf{r}_{mn}=\mathbf{r}_{m}-\mathbf{r}_{n}$ for the relative coordinate of the $m$-th and $n$-th nucleons, and $\ket{\Psi_{I(F)}(0^+)}$ the wave functions of the nuclear initial and final states with spin-parity $0^+$. The superscript $a$ represents different decay mechanisms, and $V^a_i$ represent the corresponding transition potentials in momentum space, whose specific forms are given in Sec.~\ref{sec:potentials}. 
In the following Sec.~\ref{sec:combinedNME}, the sub-amplitudes $\mathcal{A}_i$ are decomposed into linear combinations of NMEs $M^a_i$, governed by different decay mechanisms, and multiplied by a series of unknown coefficients $\mathcal{C}$s, which are related to the Wilson coefficients of LNV operators at high energy.

\subsection{Leading-order LNV transition potentials}
\label{sec:potentials}

\begin{figure}[]
 \centering
 \includegraphics[width=\columnwidth,clip=,]{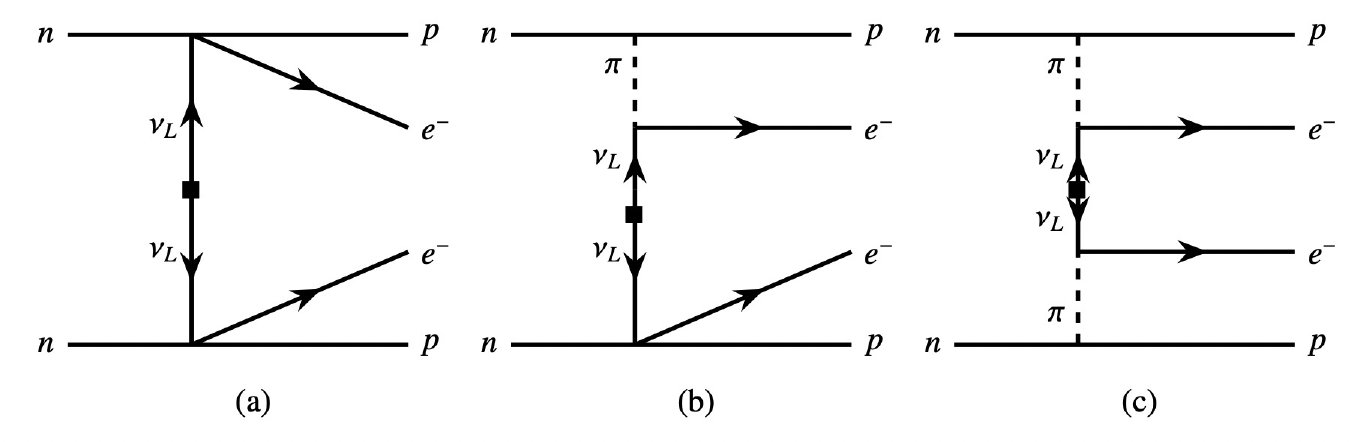}  
\caption{The Feynman diagrams in the standard mechanisms of $\onbb$ decay governed by the exchange of light Majorana neutrinos (type I). The black-square symbols represent LNV interactions throughout this paper.}
\label{fig:Majorana_neutrino} 
 \end{figure}

\begin{figure}[]
 \centering
 \includegraphics[width=\columnwidth,clip=,]{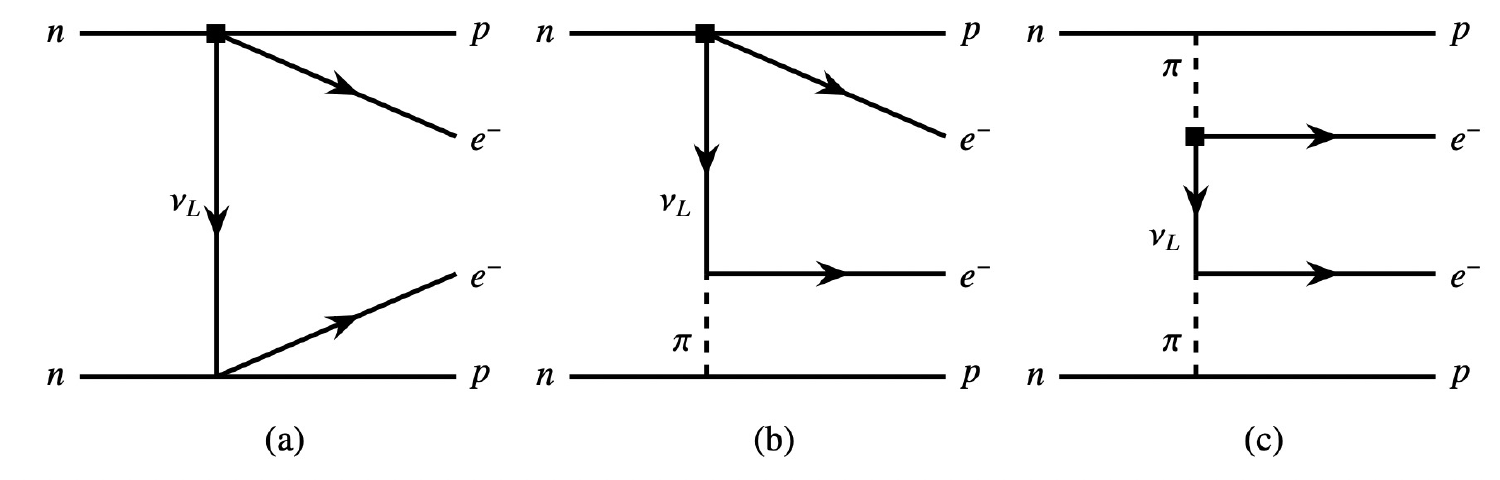}  
\caption{The Feynman diagrams in the nonstandard mechanisms of $\onbb$ decay governed by neutrino exchange without mass insertion (type II).}
\label{fig:Dirac_exchange}
\end{figure}

In this section, we present the transition potentials $V^a_i$ corresponding to three different types of mechanisms, including the standard mechanism of exchanging light Majorana neutrinos, the momentum-dependent long-range mechanism, and the short-range mechanism, which are also called type I, II, III mechanisms in this work, respectively. In each type of mechanism, depending on whether there is the exchange of zero, one, or two pions between the hadrons and leptons, the mechanism is further decomposed into $NN$, $\pi N$, and $\pi\pi$ terms.
In this study, we only consider the LO transition potentials, which can be regarded as dominant contributions.

In the type-I mechanism,  the LO transition potentials induced by the diagrams of Fig.~\ref{fig:Majorana_neutrino}(a), (b) and (c) are given by
\bsub\begin{align} 
    V^{\Rmnum{1},NN}_{L}
    &=\frac{V_{u d}^{2}}{\mathbf{q}^{2}} \frac{m_{\beta \beta}}{m_e}\left(\frac{g_{V}^{2}}{g_{A}^{2}} h^{VV}_{F}\left(\mathbf{q}^{2}\right)-\boldsymbol{\sigma}^{(1)} \cdot \boldsymbol{\sigma}^{(2)} h^{AA}_{G T}\left(\mathbf{q}^{2}\right)\right),  \\
    V_{L}^{\Rmnum{1},\pi N}&= \frac{V_{u d}^{2}}{\mathbf{q}^2}\frac{m_{\beta\beta}}{m_e} \left(-\boldsymbol{\sigma}^{(1)}\cdot\boldsymbol{\sigma}^{(2)}h^{AP}_{GT}(\mathbf{q}^2)-S^{(12)}h^{AP}_{T}(\mathbf{q}^2)\right),\\
    V_{L}^{\Rmnum{1},\pi \pi}&=  \frac{V_{u d}^{2}}{\mathbf{q}^2} \frac{m_{\beta\beta}}{m_e}\left(-\boldsymbol{\sigma}^{(1)}\cdot\boldsymbol{\sigma}^{(2)}h^{PP}_{GT}(\mathbf{q}^2)-S^{(12)}h^{PP}_{T}(\mathbf{q}^2)\right),
\end{align}
\esub
where $V_{ud}=0.97373$~\cite{PDG:2022} is an element of the Cabibbo-Kobayashi-Maskawa  matrix. The effective neutrino mass $m_{\beta\beta}$ is a linear combination of all the three neutrino masses, $m_{\beta\beta}=\sum_{j=1}^3 U^2_{ej}m_{j}$, where $U_{ej}$ are the elements of the Pontecorvo-Maki-Nakagawa-Sakata matrix and $m_{j}$ are the eigenvalues of neutrino mass states.  It is noted that in this definition, $m_{\beta\beta}$ could take the negative values, which are kept for illustration. The tensor potential $S^{(12)}=\boldsymbol{\sigma}^{(1)}\cdot\boldsymbol{\sigma}^{(2)}- 3(\boldsymbol{\sigma}^{(1)}\cdot\mathbf{q})(\boldsymbol{\sigma}^{(2)}\cdot\mathbf{q})/\mathbf{q}^2$.
It has been found in the recent studies~\cite{Cirigliano:2018PRL,Cirigliano:2019PRC}  for the $nn\to ppe^-e^-$ process based on the type-I mechanism that a contact transition potential
\beq 
 V_{L}^{I,CT} =  V_{u d}^{2}  \frac{m_{\beta \beta}}{m_e}(-2g^{NN}_\nu) \frac{1}{g^2_A}h_F^{VV}(\mathbf{q}^{2})
\eeq 
is required to be included at the chiral LO to ensure renormalizability. However, the low-energy constant $g^{NN}_\nu$ of this term is scheme and scale dependent~\cite{Wirth:2021PRL}. Whether the contact transition operator needs to be introduced in phenomenological nuclear models remains an open question. Moreover, a recent study demonstrated that in the relativistic framework, the transition amplitude of the $nn \to ppe^-e^-$ process is renormalized without promoting the contact neutrino potential term to the LO~\cite{Yang:2023}. In light of these facts, we do not consider the contribution of this term throughout this work.

The neutrino potentials in momentum space $h^K_{\alpha}(\mathbf{q}^2)$  with $\alpha\in\{F,GT,T\}$ and $K\in\{VV,AA,AP,PP\}$ are defined as
\begin{equation}
\label{eq:neutrino_potentials}
    \begin{aligned}
        h^{VV}_{F}\left(\mathbf{q}^{2}\right)= \frac{g_V^2(\mathbf{q}^2)}{g_V^2},&\quad h_{G T}^{A A}\left(\mathbf{q}^{2}\right)  =\frac{g_{A}^{2}(\mathbf{q}^{2})}{g_{A}^{2}}, \\
        h_{G T}^{A P}\left(\mathbf{q}^{2}\right)
        =\frac{ g_{A}(\mathbf{q}^{2})g_{P}(\mathbf{q}^{2})}{g_{A}^{2}} \frac{\mathbf{q}^{2}}{3 m_{N}},&\quad h_{G T}^{P P}\left(\mathbf{q}^{2}\right)=\frac{g_{P}^{2}(\mathbf{q}^{2})}{g_{A}^{2}} \frac{\mathbf{q}^{4}}{12 m_{N}^{2}},\\
        h_T^{AP}\left(\mathbf{q}^{2}\right) =-h_{GT}^{AP}\left(\mathbf{q}^{2}\right),&\quad h_T^{PP}\left(\mathbf{q}^{2}\right)  =-h_{GT}^{PP}\left(\mathbf{q}^{2}\right).
    \end{aligned}
\end{equation}
At LO the form factors read~\cite{Engel:2017}
\begin{align}
    g_V(\mathbf{q}^2)&=g_V=1,\quad g_A(\mathbf{q}^2)=g_A\simeq 1.27,\nonumber\\
    g_P(\mathbf{q}^2)&=-g_A\frac{2m_N}{\mathbf{q}^2+m^2_{\pi}}.
\end{align}

In the type-II mechanism, the LO transition potentials induced by the diagrams of Fig.~\ref{fig:Dirac_exchange}(a), (b) and (c) are  expressed as
\begin{subequations}
    \begin{align}
        V^{\Rmnum{2},NN}_{E}=& \frac{V_{ud}}{\mathbf{q}^2}\Bigg[\mathcal{C}^{\Rmnum{2}}_{1} \left( -\frac{1}{3}\frac{g_V^2}{g_A^2}h^{VV}_F(\mathbf{q}^2)\right) \nonumber\\
        &+\mathcal{C}^{\Rmnum{2}}_{2}\left(-\frac{2}{9}\boldsymbol{\sigma}^{(1)}\cdot\boldsymbol{\sigma}^{(2)}h^{AA}_{GT}(\mathbf{q}^2)-\frac{1}{9}S^{(12)}h^{AA}_T(\mathbf{q}^2)\right)\Bigg],\\ 
        V^{\Rmnum{2},NN}_{m_e}=&\frac{ V_{ud}}{\mathbf{q}^2}\Bigg[{\mathcal{C}^{\Rmnum{2}}_{1}\left(\frac{1}{6}\frac{g_V^2}{g_A^2}h^{VV}_F(\mathbf{q}^2)\right)} \nonumber\\
        &+\mathcal{C}^{\Rmnum{2}}_{2}\left(-\frac{1}{18}\boldsymbol{\sigma}^{(1)}\cdot\boldsymbol{\sigma}^{(2)}h^{AA}_{GT}(\mathbf{q}^2)+\frac{2}{9}S^{(12)}h^{AA}_T(\mathbf{q}^2)\right)\Bigg],\\        \label{eq:LRpiNpotential}
        V^{\Rmnum{2},\pi N}_{L}=&  \frac{ V_{ud}}{\mathbf{q}^2} \Bigg[\mathcal{C}^{\Rmnum{2}}_{3}\left(\frac{1}{2}\boldsymbol{\sigma}^{(1)}\cdot\boldsymbol{\sigma}^{(2)}h^{AP}_{GT}(\mathbf{q}^2)+\frac{1}{2}S^{(12)}h^{AP}_T(\mathbf{q}^2)\right)\Bigg],\\
        V^{\Rmnum{2},\pi N}_{m_e}=&\frac{ V_{ud}}{\mathbf{q}^2}\Bigg[\mathcal{C}^{\Rmnum{2}}_{2}\left(-\frac{1}{2}\boldsymbol{\sigma}^{(1)}\cdot\boldsymbol{\sigma}^{(2)}h^{AP}_{GT}(\mathbf{q}^2)-\frac{1}{2}S^{(12)}h^{AP}_T(\mathbf{q}^2)\right)\Bigg],\\ 
        \label{eq:LRpipipotential}
        V^{\Rmnum{2},\pi \pi}_{L}=&\frac{V_{ud}}{\mathbf{q}^2}\Bigg[\mathcal{C}^{\Rmnum{2}}_{3}\left(\boldsymbol{\sigma}^{(1)}\cdot\boldsymbol{\sigma}^{(2)}h^{PP}_{GT}(\mathbf{q}^2)+S^{(12)}h^{PP}_T(\mathbf{q}^2)\right)\Bigg],\\
        V^{\Rmnum{2},\pi \pi}_{m_e}=&\frac{V_{ud}}{\mathbf{q}^2}\Bigg[\mathcal{C}^{\Rmnum{2}}_{2}\left(-\frac{1}{2}\boldsymbol{\sigma}^{(1)}\cdot\boldsymbol{\sigma}^{(2)}h^{PP}_{GT}(\mathbf{q}^2)-\frac{1}{2}S^{(12)}h^{PP}_T(\mathbf{q}^2)\right)\Bigg]\;,
    \end{align}
\end{subequations} 
where the unknown coefficients $C$s are the low-energy constants (LECs) for the transition potentials, and they are related to the Wilson coefficients~\cite{Cirigliano:2017}. The corresponding relations are given in the Appendix.

\begin{figure}[]
 \centering
 \includegraphics[width=\columnwidth,clip=,]{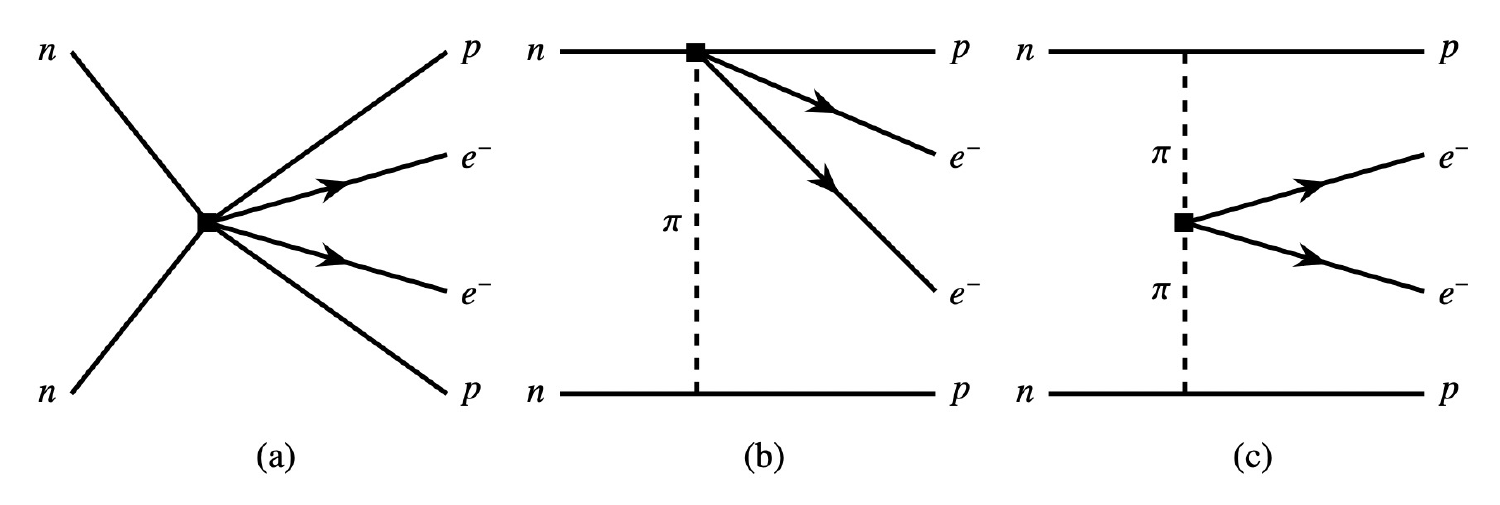}  
\caption{The Feynman diagrams in the nonstandard mechanisms  of $\onbb$ decay governed by the short-range contact terms  (type III).}
\label{fig:Short_range_contact}
\end{figure}

In the final part of this subsection, we calculate the transition potentials of the diagrams which are induced by short-range contact terms, as depicted in Fig.~\ref{fig:Short_range_contact}. In physics  beyond the standard model, the contributions of these contact terms might arise from the exchange of heavy neutrinos, and the degrees of freedom of these heavy neutrinos do not manifest in the low-energy processes described by chiral EFT.

In momentum space, the LO transition potential induced by the diagrams of Fig.~\ref{fig:Short_range_contact}(a), (b) and (c) are, respectively, given by
\begin{subequations}
    \begin{align}
        V^{\Rmnum{3},NN}_{L(R,M)}=&\frac{1}{m^2_{\pi}} \mathcal{C}^{\Rmnum{3}}_{1L(R,M)}\left(-\frac{2g_V^2}{g_A^2}h^{VV}_F(\mathbf{q}^2)\right),\\
        V^{\Rmnum{3},\pi N}_{L(R,M)}=& \frac{1}{m^2_{\pi}} \mathcal{C}^{\Rmnum{3}}_{2L(R,M)}\left(\frac{1}{2}\boldsymbol{\sigma}^{(1)}\cdot\boldsymbol{\sigma}^{(2)}h^{AP}_{GT}(\mathbf{q}^2)+\frac{1}{2}S^{(12)}h^{AP}_T(\mathbf{q}^2)\right),\\ 
  V^{\Rmnum{3},\pi\pi}_{L(R)}=&\frac{1}{m^2_{\pi}} \mathcal{C}^{\Rmnum{3}}_{3 L(R)}\Bigg[ \left(-\frac{1}{2}h^{AP}_{GT}(\mathbf{q}^2)-h^{PP}_{GT}(\mathbf{q}^2)\right)\boldsymbol{\sigma}^{(1)}\cdot\boldsymbol{\sigma}^{(2)} \nonumber\\ & +\left(-\frac{1}{2}h^{AP}_{T}(\mathbf{q}^2)-h^{PP}_{T}(\mathbf{q}^2)\right)S^{(12)}\Bigg].
       \end{align}
    \end{subequations}

\subsection{NMEs in different mechanisms}
\label{sec:combinedNME}
According to Eq.(\ref{eq:subamplitude}), the sub-amplitudes $\mathcal{A}_i$ can be expressed as linear combinations of the combined NMEs $\mathcal{M}^a_i$ multiplied by the effective neutrino mass $m_{\beta\beta}/m_e$ or the unknown coefficients $\mathcal{C}$s,
\begin{subequations} 
\label{eq:amplitudeNME}
 \begin{align}
    \mathcal{A}_L &= \dfrac{m_{\beta\beta}}{m_e} \left(\mathcal{M}_L^{I,NN} + \mathcal{M}_L^{I,\pi N} + \mathcal{M}_L^{I,\pi\pi} \right) \\
    &\quad +  \mathcal{C}_3^{II} \left( \mathcal{M}_{L}^{II,\pi N} + \mathcal{M}_{L}^{II,\pi\pi} \right)\nn\\
    &\quad + \mathcal{C}_{1L}^{III} \mathcal{M}_L^{III,NN} +  \mathcal{C}_{2L}^{III} \mathcal{M}_L^{III,\pi N} +  \mathcal{C}_{3L}^{III} \mathcal{M}_L^{III,\pi \pi}\;,\\
    \mathcal{A}_R &= \mathcal{C}^{III}_{1R} \mathcal{M}_R^{III,NN} +  \mathcal{C}^{III}_{2R} \mathcal{M}_R^{III,\pi N} +  \mathcal{C}^{III}_{3R} \mathcal{M}_R^{III,\pi\pi} \;,\\
    \mathcal{A}_E &=\mathcal{C}^{II}_{1}\mathcal{M}^{II,NN}_{E,1}+\mathcal{C}^{II}_{2}\mathcal{M}^{II,NN}_{E,2}\;,\\
    \mathcal{A}_{m_e} &=\mathcal{C}^{II}_{1}\mathcal{M}^{II,NN}_{m_e,1}+\mathcal{C}^{II}_{2}\left(\mathcal{M}^{II,NN}_{m_e,2}+\mathcal{M}^{II,\pi N}_{m_e}+\mathcal{M}^{II,\pi \pi}_{m_e}\right)\;,\\
    \mathcal{A}_M &= \mathcal{C}^{III}_{1M} \mathcal{M}_M^{III,NN}+\mathcal{C}^{III}_{2M} \mathcal{M}_M^{III,\pi N}\;.
 \end{align}
\end{subequations}
The NMEs $\mathcal{M}^a_i$ are combinations of Fermi (F), Gamow-Teller (GT) and Tensor (T)-types of NMEs $M^K_{\alpha, (sd)}$. For those corresponding to the type-I mechanism,  one finds
\begin{subequations}
\label{eq:combineNME-type1} 
 \begin{align}
     {\mathcal M}^{\Rmnum{1},NN}_{L}
    &=V^2_{ud}\left(\frac{g^2_V}{g^2_A}M_F^{VV}-M^{AA}_{GT}\right),\\ \mathcal{M}^{\Rmnum{1},\pi N}_{L}
    &=-V^2_{ud}\left(M^{AP}_{GT}+M^{AP}_T\right),\\
    \mathcal{M}^{\Rmnum{1},\pi\pi}_{L}
    &=-V^2_{ud}\left(M^{PP}_{GT}+M^{PP}_T\right), 
 \end{align}
\end{subequations}
  those of the type-II mechanism
\begin{subequations}
\label{eq:combineNME-type2} 
 \begin{align}
    \mathcal{M}^{\Rmnum{2},NN}_{E,1}
    &=-\frac{V_{ud}}{3}\frac{g_V^2}{g_A^2}M_F^{VV},\\
    \mathcal{M}^{\Rmnum{2},NN}_{E,2}
    &=-V_{ud}\left(\frac{2}{9}M^{AA}_{GT}+\frac{1}{9}M^{AA}_T\right),\\
    \mathcal{M}^{\Rmnum{2},NN}_{m_e,1}
    &=-\frac{1}{2}\mathcal{M}^{\Rmnum{2},NN}_{E,1},\\
    \mathcal{M}^{\Rmnum{2},NN}_{m_e,2}
    &=-V_{ud}\left(\frac{1}{18}M^{AA}_{GT}-\frac{2}{9}M^{AA}_T\right),\\
    \mathcal{M}^{\Rmnum{2},\pi N}_{L}
    &=V_{ud}\left(\frac{1}{2}M^{AP}_{GT}+\frac{1}{2}M^{AP}_T\right),\\ 
    \mathcal{M}^{\Rmnum{2},\pi N}_{m_e}&=-\mathcal{M}^{\Rmnum{2},\pi N}_{L},\\
\mathcal{M}^{\Rmnum{2},\pi\pi}_{L}
&=V_{ud}\left(M^{PP}_{GT}+M^{PP}_T\right),\\ \mathcal{M}^{\Rmnum{2},\pi\pi}_{m_e}
&=-\frac{1}{2}\mathcal{M}^{\Rmnum{2},\pi\pi}_{L},
 \end{align}
\end{subequations}
and  those of the type-III mechanism
\begin{subequations}
\label{eq:combineNME-type3} 
 \begin{align}
    \mathcal{M}^{\Rmnum{3},NN}_{L(R,M)}
    &=-\frac{2g^2_V}{g^2_A}M^{VV}_{F,sd},\\
    \mathcal{M}^{\Rmnum{3},\pi N}_{L(R,M)}
    &=\frac{1}{2}M^{AP}_{GT,sd}+\frac{1}{2}M^{AP}_{T,sd},\\
    \mathcal{M}^{\Rmnum{3},\pi\pi}_{L(R)}&=-\frac{1}{2}M^{AP}_{GT,sd}
    -\frac{1}{2}M^{AP}_{T,sd}-M^{PP}_{GT,sd}-M^{PP}_{T,sd}.
 \end{align}
\end{subequations}
 On the right-hand side of the above equations, each component of the NMEs, namely $M^K_{\alpha, (sd)}$, is computed with the corresponding transition operator and nuclear wave functions, 
\begin{equation}
    \label{eq:component_NME}
    \begin{aligned}
        M^{K}_{\alpha, (sd)} &=\bra{\Psi_F(0^{+})}\mathcal{O}^K_{\alpha, (sd)}\ket{\Psi_I(0^{+})},
    \end{aligned}
\end{equation}
with $\alpha\in\{F, GT, T\}$, and $K\in\{VV,AA,AP,PP\}$. The transition operators are defined as 

\begin{equation}
    \begin{aligned}
        \mathcal{O}^K_{\alpha} 
        &=\frac{2R_A}{\pi}\sum_{m,n}\int dq h^K_{\alpha}(\mathbf{q}^2)j_\lambda(qr_{mn})\tau^{(m)+}\tau^{(n)+},\\  \mathcal{O}^K_{\alpha, sd} 
        &=\frac{2R_A}{\pi m_\pi^2}\sum_{m,n}\int dq q^2 h^K_{\alpha, sd}(\mathbf{q}^2)j_\lambda(qr_{mn})\tau^{(m)+}\tau^{(n)+},\\
    \end{aligned}
\end{equation}
where $q=|\mathbf{q}|$, the rank of the spherical Bessel function $j_\lambda(qr_{mn})$ is $\lambda=0$ for the Fermi and GT terms, and $\lambda=2$ for the tensor terms.

\begin{figure*}[h!]
 \centering
 \includegraphics[width=2\columnwidth]{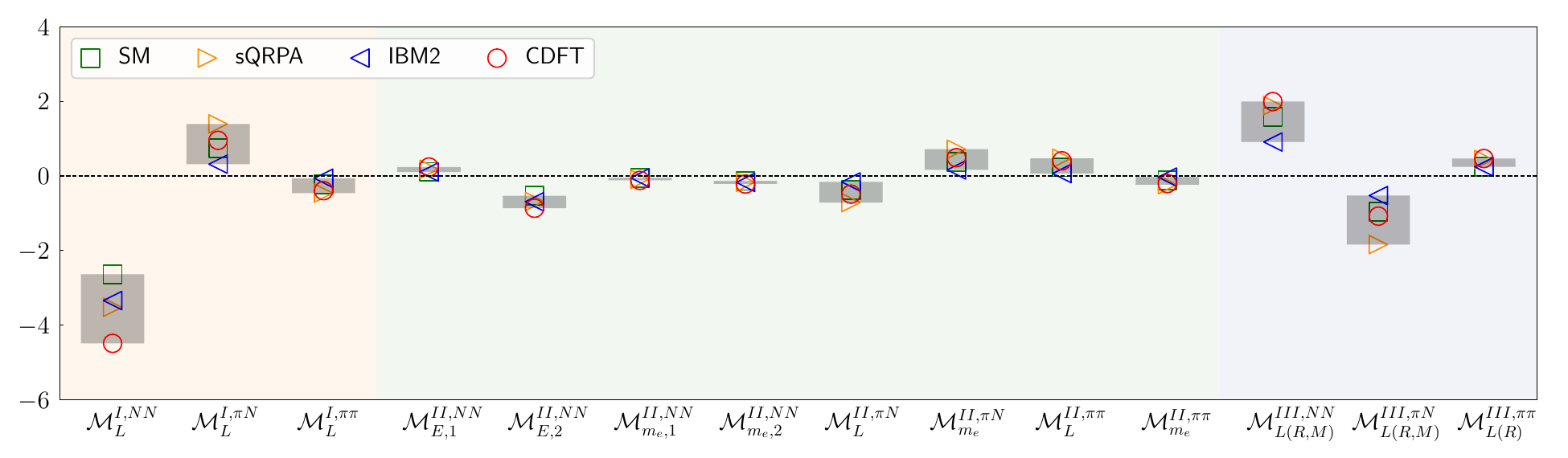}  
\caption{The combined NMEs of $0\nu\beta\beta$ decay in \nuc{Xe}{136} from the MR-CDFT calculation, in comparison with the results of other nuclear models, including SM~\cite{Menendez:2018JPG}, sQRPA~\cite{Hyvarinen:2015}, IBM2~\cite{Deppisch:2020ztt}. The orange background region contains the NMEs for type I decay mechanism, while the green and blue regions are for type II and type III mechanisms, respectively.
}
\label{fig:NMEcompare}
\end{figure*}

\section{Results and discussion}
\label{sec:results}
\subsection{The NMEs from MR-CDFT calculations}

The wave functions $\ket{\Psi_{I(F)}(0^{+})}$ of the ground states of initial and final nuclei in the $0\nu\beta\beta$ decay are obtained from the MR-CDFT~\cite{Yao:2010,Yao:2014PRC} calculation based on the relativistic point-coupling density functional PC-PK1~\cite{Zhao:2010PRC}, where the wave functions are given by a linear combination of both angular-momentum and particle-number projected axially-deformed mean-field states. The Dirac equation for nucleons in each mean-field state is solved in a spherical harmonic oscillator basis within $10$ major shells.  Pairing correlations between nucleons are treated by the BCS method using a zero-range pairing force with the strength parameters chosen as $-314.550$ MeV fm$^3$ and $-346.500$ MeV fm$^3$ for neutrons and protons, respectively. See Refs.~\cite{Song:2014,Yao:2015,Yao:2016PRC,Song:2017,Ding:2023} for more details.

\begin{table}[bt]
    \centering  \renewcommand{\arraystretch}{1.2}
    \caption{The NMEs $M^K_{\alpha,(sd)}$ of $0\nu\beta\beta$ decay, defined in (\ref{eq:component_NME}), in different candidate nuclei from the MR-CDFT calculations.}  
 	\label{tab:CDFTNMEs} 
    \begin{tabular}{crrrrr}
    \toprule
        NME & $\nuc{Ge}{76}$ & $\nuc{Se}{82}$ & $\nuc{Mo}{100}$ & $\nuc{Te}{130}$ & $\nuc{Xe}{136}$ \\ \midrule
        $M_F^{VV}$ & $-1.924$ & $-1.742$ & $-1.814$ & $-1.365$ & $-1.184$ \\ 
         \hline
        $M_{GT}^{AA}$ & $5.743$ & $5.021$ & $6.341$ & $4.565$ & $4.003$ \\ 
        $M_{GT}^{AP}$ & $-1.462$ & $-1.320$ & $-1.591$ & $-1.206$ & $-1.059$\\ 
        $M_{GT}^{PP}$ & $0.423$ & $0.386$ & $0.461$ & $0.352$ & $0.308$ \\ 
        $M_{GT}^{MM}$ & $0.326$ & $0.298$ & $0.359$ & $0.275$ & $0.240$ \\ 
         \hline
        $M_{T}^{AA}$ & $-$ & $-$ & $-$ & $-$ & $-$ \\ 
        $M_{T}^{AP}$ & $0.020$ & $-0.018$ & $0.039$ & $0.062$ & $0.053$ \\ 
        $M_{T}^{PP}$ & $0.174$ & $0.156$ & $0.162$ & $0.121$ & $0.108$ \\ 
        $M_{T}^{MM}$ & $0.012$ & $0.005$ & $0.012$ & $0.017$ & $0.014$ \\ 
         \hline
        $M_{F,sd}^{VV}$ & $-2.226$ & $-2.049$ & $-2.484$ & $-1.843$ & $-1.607$ \\ 
        $M_{GT,sd}^{AA}$ & $7.141$ & $6.549$ & $7.974$ & $6.160$ & $5.390$ \\ 
        $M_{GT,sd}^{AP}$ & $-3.120$ & $-2.882$ & $-3.513$ & $-2.728$ & $-2.379$ \\ 
        $M_{GT,sd}^{PP}$ & $1.090$ & $1.011$ & $1.239$ & $0.966$ & $0.841$ \\ 
        $M_{T,sd}^{AP}$ & $0.143$ & $0.023$ & $0.084$ & $0.271$ & $0.221$ \\ 
        $M_{T,sd}^{PP}$ & $-0.226$ & $-0.196$ & $-0.284$ & $-0.270$ & $-0.227$ \\ \bottomrule
    \end{tabular}
\end{table}

\begin{figure}[]
 \centering
 \includegraphics[width=\columnwidth,clip=,]{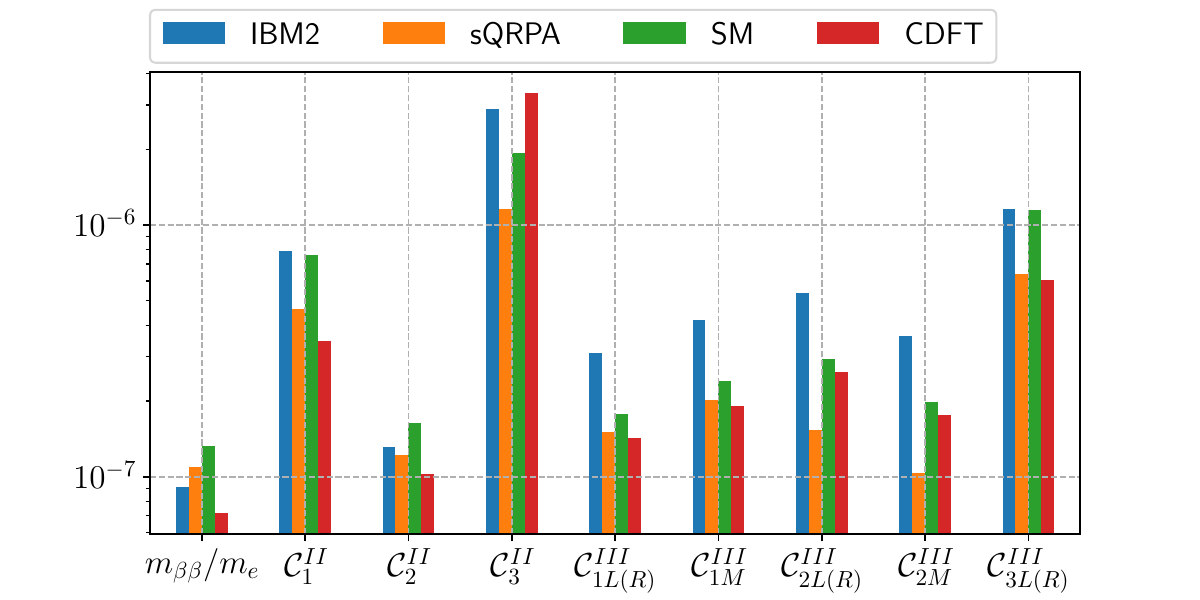}  
\caption{Comparison of the upper limits of the dimensionless unknown coefficients $m_{\beta\beta}/m_e$ and ${\cal C}$s, derived on the basis of the half-life limit of $^{136}$Xe from KamLAND-Zen~\cite{Abe:2023PRL} and the combined NMEs ${\cal M}^a_i$ from different nuclear models. Here, we assume that all of the coefficients are positive, and the $0\nu\beta\beta$ decay is driven by the mechanisms depending on only one of these  unknown coefficients.
}
\label{fig:1Dconstraint}
\end{figure}

Table~\ref{tab:CDFTNMEs} lists the values of the NMEs $M^{K}_{\alpha, (sd)}$ of the five popular candidate nuclei  $^{76}$Ge, $^{82}$Se,  $^{100}$Mo, $^{130}$Te, and $^{136}$Xe.
According to Eqs.(\ref{eq:combineNME-type1})-(\ref{eq:combineNME-type3}), the combinations of these NMEs $M^{K}_{\alpha, (sd)}$ define the combined NMEs $\mathcal{M}^a_i$, which, together with the best lower limits of $0\nu\beta\beta$-decay half-life, could be used to constrain the unknown coefficients ${\cal C}$s of the LNV operators. Such kind of studies have been carried out based on a specific neutrino mass model using the NMEs from different  nuclear models~\cite{Dekens:2023PRC,Deppisch:2020ztt,Kotila:2021longrange,Chen:2024,Graf:2022PRD,Deppisch:2007prl,Horoi:2017gmj}. Our NMEs provide a complementary choice for these studies, and are helpful for quantifying the uncertainty in the NMEs from nuclear model calculations. Reducing the discrepancy is difficult because each phenomenological model has its own assumptions and uncontrolled approximations. In recent years, advances in the development of ab initio methods have enabled the calculations of NMEs for $0\nu\beta\beta$ decay~\cite{Yao:2020PRL,Belley:2021PRL,Novario:2021PRL,Belley2023TeXe}.  However, all of these studies are still limited to the type-I mechanism.

It has been shown in Ref.~\cite{Cirigliano:2018JHEP} that certain component $M^K_{\alpha,(sd)}$ of the NMEs, such as  $M^{AP}_{T, sd}$ related to the nonstandard mechanisms may vary by more than one order of magnitude with nuclear models. In contrast,  the discrepancy in the combined NMEs among different models is much smaller than that appearing in $M^K_{\alpha,(sd)}$. This justifies the validity of combining the NMEs based on the diagrams in Figs.~\ref{fig:Majorana_neutrino}, \ref{fig:Dirac_exchange}, and \ref{fig:Short_range_contact}.  As shown in Fig.~\ref{fig:NMEcompare},  a variation of approximately a factor of $2-4$ is shown in most of the combined NMEs ${\cal M}^a_i$ for \nuc{Xe}{136} produced by different models, including the interacting SM~\cite{Menendez:2017JPCS}, IBM2~\cite{Deppisch:2020ztt}, spherical QRPA (sQRPA)~\cite{Hyvarinen:2015}, based on both standard and nonstandard mechanisms. It indicates that the actual impact of the uncertainty from nuclear models  on the constraints of new physics in the nonstandard mechanisms is comparable to that in the standard mechanism.  It is shown in Fig.~\ref{fig:NMEcompare} that in general the sizes of the NMEs from the type-I and type-III mechanisms are overall larger than those from the type-II mechanism, defined in (\ref{eq:combineNME-type2}).

\subsection{Constraints on the  coefficients of LNV operators with $^{136}$Xe }
\label{subsec:constraints}

\begin{figure}[bt]
 \centering
 \includegraphics[width=\columnwidth]{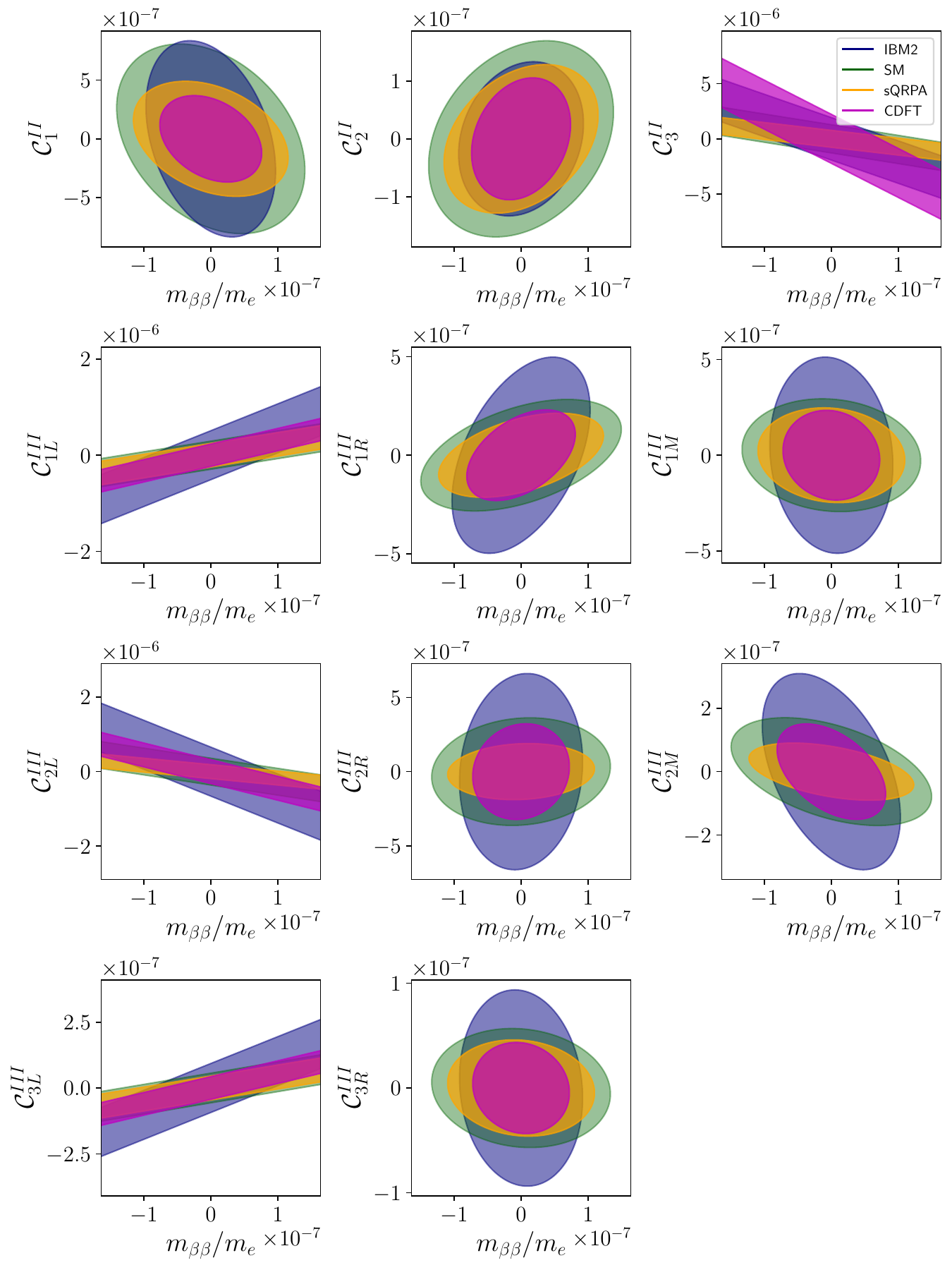}  
\caption{Same as Fig.~\ref{fig:1Dconstraint}, but the $0\nu\beta\beta$ decay is assumed to be driven by the standard mechanism, together with nonstandard mechanisms depending on one of the  unknown coefficients ${\cal C}$s. 
}
\label{fig:2Dconstraint}
\end{figure}

\nuclide[136]{Xe} is currently the candidate nucleus of $0\nu\beta\beta$ decay with the best half-life sensitivity of $T^{0\nu}_{1/2}>2.3\times 10^{26}$ years at 90\% confidence level~\cite{KamLAND-Zen:2022}. 
In the following, we will discuss the constraints on the unknown coefficients $\mathcal{C}$s of LNV potentials based on the combined NMEs in Fig.~\ref{fig:NMEcompare}, the half-life limit, and the phase-space factors $G_{0i}$ for \nuc{Xe}{136}~\cite{Scholer:2023},
\begin{equation}
    \begin{aligned}
        G_{01}&=2.09,\quad G_{02}=5.15,\quad G_{03}=1.40,\\
        G_{04}&=1.88,\quad G_{06}=2.86, \quad G_{09}=4.59,
    \end{aligned}
\end{equation}
which are in units of $10^{-14}$ yr$^{-1}$. This kind of analysis is carried out based on the $\nu$DoBe program~\cite{Scholer:2023}. We note that this program has already been employed  to unravel different mechanisms of $0\nu\beta\beta$ decay using the NMEs computed with the IBM2 nuclear model~\cite{Graf:2022PRD}. Considering the fact that  some LECs are unknown, in our work, we combine the Wilson coefficients of LNV operators and LECs into a new set of independent $\mathcal{C}$s (see the appendix for details).
 
Figure~\ref{fig:1Dconstraint} presents the upper limits of the coefficients $\mathcal{C}$, which are derived from the half-life and the NMEs depending on only one of the $\mathcal{C}$s. For comparison, the combined NMEs from different nuclear models are used. One can observe that for the NMEs from all nuclear models, the constraint from the half life of \nuclide[136]{Xe} on $m_{\beta\beta}/m_e$ is the most stringent,  while that on  $\mathcal{C}^{II}_3$ is weakest among the 12 coefficients. 
From Eqs. (\ref{eq:LRpiNpotential}) and (\ref{eq:LRpipipotential}), one finds that the coefficient $\mathcal{C}^{II}_3$ corresponds to two LNV potentials: $V^{II,\pi N}_L$ and $V^{II,\pi \pi}_L$. If the $0\nu\beta\beta$ decay is solely driven by these neutrino potentials, the half-life will depend on the sum of the combined NMEs $\mathcal{M}^{II,\pi N}_{L}$ and $\mathcal{M}^{II,\pi \pi}_{L}$. It can be seen from Fig.~\ref{fig:NMEcompare} that these two NMEs cancel each other, as also found in Ref.~\cite{Agostini:2022RMP}. Consequently, the half-life is not sensitive to $\mathcal{C}^{II}_3$. Nevertheless, the upper limits of the coefficients ${\cal C}$s  are generally in the same order of magnitude as $m_{\beta\beta}/m_e$. Compared to the results based on the NMEs by other nuclear models, the use of the NME by CDFT provides the strongest constraint on the effective neutrino mass $m_{\beta\beta}$, and comparable constraints on other 11 coefficients.

\subsection{Correlations of LNV coefficients}

If the $0\nu\beta\beta$ decay is driven not only by the standard mechanism but also by nonstandard mechanisms that are only related to one of the 11 coefficients, one needs to consider the correlations of LNV coefficients. In Eq.~(\ref{eq:half-life-2}), there are interference terms: (1) among different contributions to $\mathcal{A}_L$ and (2) between $\mathcal{A}_L$ and the other sub-amplitudes. Consequently, constraints on the LNV coefficients from the interpretation of the half-life of $\onbb$ decay limit would exhibit linear or elliptical correlation, respectively. Fig.~\ref{fig:2Dconstraint} displays two-dimensional plots of $m_{\beta\beta}/m_e$ and the other LNV coefficients for the NMEs from different nuclear models.

The LNV transition potentials with the subscript $L$ contribute to the sub-amplitude $\mathcal{A}_L$, accompanied by the same leptonic structure.
Thus the allowed region of $m_{\beta\beta}/m_e$ and each of the LNV coefficients  $C^{II}_3$, and $C^{III}_{iL}$ $(i=1,2, 3)$ will fall into bands, as shown in Fig.~\ref{fig:2Dconstraint}. In this case, one can determine rather precisely the value of $C^{II}_3$, or $C^{III}_{iL}$ at the time when the half-life and effective neutrino mass are known. 
In contrast, the LNV potentials with the subscript $m_e$, $E$, $R$ or $M$ contribute to the sub-amplitudes $\mathcal A_{m_e}$, $\mathcal A_E$, $\mathcal A_R$ or $\mathcal A_M$, respectively. Therefore, the allowed region of $m_{\beta\beta}/m_e$ and each of the LNV coefficients $\mathcal{C}^{II}_{1(2)}$, $\mathcal{C}^{III}_{iR}$ $(i=1,2,3)$ and $\mathcal{C}^{III}_{1(2)M}$ lies within ellipse. 

From Fig.~\ref{fig:NMEcompare}, the combined NMEs $\mathcal{M}_L^{I,NN} + \mathcal{M}_L^{I,\pi N} + \mathcal{M}_L^{I,\pi\pi}$,  $\mathcal{M}_L^{II,\pi N} + \mathcal{M}_L^{II,\pi \pi}$, $ \mathcal{M}_{L(R,M)}^{III,\pi N}$ are negative, while $\mathcal{M}_{L(R,M)}^{III,NN}$ and  $\mathcal{M}_{L(R)}^{III,\pi\pi}$ are positive. Thus, according to Eq.(\ref{eq:amplitudeNME}),  $m_{\beta\beta}/m_e$ is positively correlated with the LNV coefficients  $\mathcal C_{1L}^{III}$, $\mathcal C_{3L}^{III}$,  while negatively correlated with $\mathcal C_{3}^{II}$ and $\mathcal C_{2L}^{III}$. Likewise, the correlations of $m_{\beta\beta}/m_e$ with the other LNV coefficients can be easily tracked.

In contrast, the coefficients $C_1^{II}$ and $C_2^{II}$ contribute to the sub-amplitudes $\mathcal A_{m_e}$ and $\mathcal A_E$  with two different leptonic structures, both of which interfere with the sub-amplitude ${\cal A}_L$ as shown in (\ref{eq:half-life-2}). It turns out that $C^{II}_1$ is negatively correlated with $m_{\beta\beta}/m_e$, while $C^{II}_2$ is the opposite. Besides, the constraint on $C^{II}_1$ is weaker than that on $C^{II}_2$, showing that the combined NME related to $C^{II}_1$ has a smaller magnitude.

It is shown in Fig.~\ref{fig:2Dconstraint} that the parameter spaces defined by the NMEs from different nuclear models are somewhat different. In general, the spaces defined by those of CDFT are smaller than others, attributed to the larger magnitude of the predicted NMEs.

\section{Summary}
\label{sec:summary}

In this work, we have expressed the leading-order LNV transition potentials for $0\nu\beta\beta$ decay from different mechanisms in terms of a set of dimensionless unknown coefficients $\mathcal{C}$s guided by chiral effective field theory. Based on these transition potentials, we have computed corresponding nuclear matrix elements (NMEs) for \nuc{Ge}{76}, \nuc{Se}{82}, \nuc{Mo}{100}, \nuc{Te}{130} and \nuc{Xe}{136} using nuclear wave functions from the multireference covariant density functional theory (MR-CDFT) calculation. The results have demonstrated that the combined NMEs based on the Feynman diagrams at the hadronic level for the nonstandard mechanisms leads to the uncertainty by different nuclear models comparable to that for the standard mechanism. Our NMEs, complementary with others, serve important nuclear inputs to constrain the unknown parameters in some particular particle physics models for LNV processes~\cite{Horoi:2018PRC,Graf:2022PRD}. As an example,  we have used our NMEs for \nuc{Xe}{136}, together with the lower limit of the half-life of $0\nu\beta\beta$ decay, and phase-space factors, to provide upper limits of  $m_{\beta\beta}$ and $\mathcal{C}$s with the assumption that the process is driven by the transition potentials related to one or two of these coefficients. The results have shown that the use of NMEs  by different nuclear models does not dramatically change the intervals of the parameter spaces for the coefficients, even though the CDFT leads to the most stringent constraint.  It is worth pointing out that our NMEs could also be used to perform a more comprehensive analysis with multiple isotopes~\cite{Agostini:2023,Chen:2024} to unravel different mechanisms, which is beyond the scope of this work. Moreover, the expressions for the leading-order transition potentials of $0\nu\beta\beta$ decay from different mechanisms can also be used in ab initio studies.

\section*{Acknowledgments} 
 
We thank  J. Engel, C. F. Jiao and M. J. Ramsey-Musolf for fruitful discussions.  This work is partly supported by the National Natural Science Foundation of China (Grant Nos. 12375119, 12141501, and 12347105), the Guangdong Basic and Applied Basic Research Foundation (2023A1515010936, 2024A1515012668),  and the Fundamental Research Funds for the Central Universities, Sun Yat-sen University (23qnpy62).

\section*{Appendix: Relations between the coefficients $\mathcal{C}$s and Wilson coefficients}

\begin{figure*}[]
 \centering
 \includegraphics[width=2\columnwidth,clip=,]{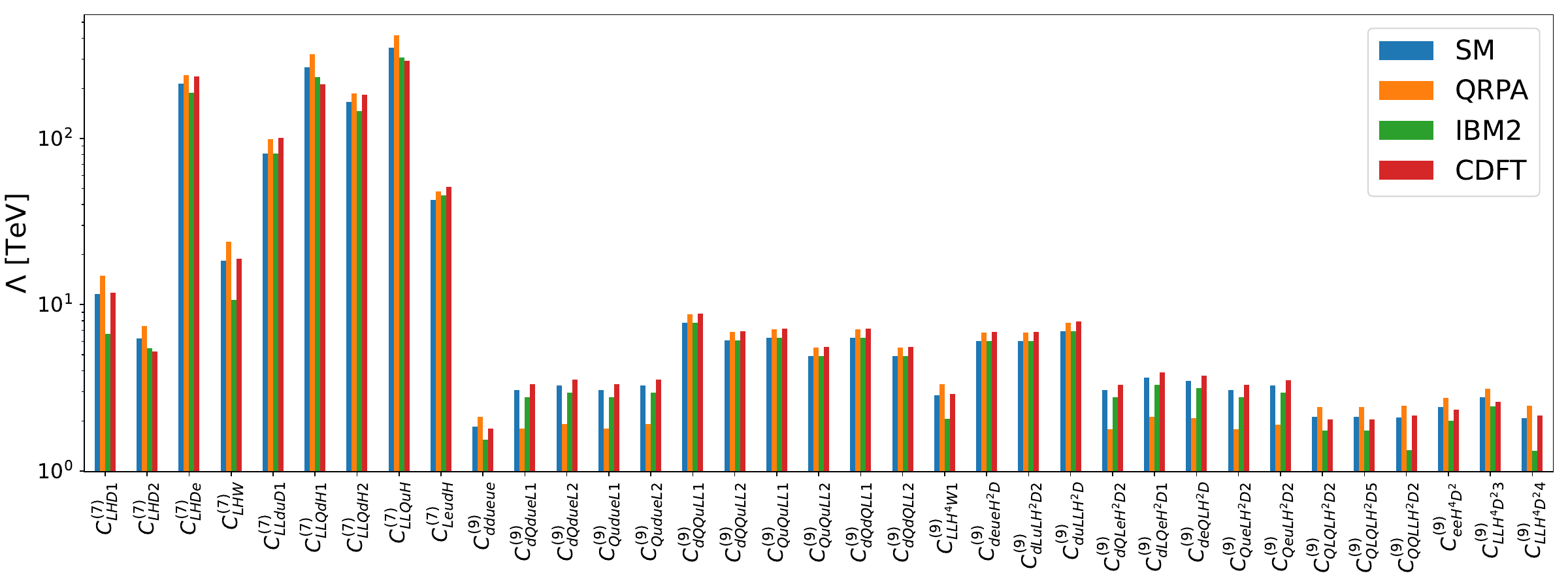}  
\caption{Comparison of the lower limits on the new physics scales for different LNV operators in the standard model EFT using the NMEs by the four different nuclear models. The notations of the Wilson coefficients of the LNV operators follow Ref.~\cite{Scholer:2023}. }
\label{fig:ScaleLimits}
\end{figure*}

In this appendix we compare the unknown coefficients $\mathcal{C}$s defined in this work to the Wilson coefficients previously used in Refs.~\cite{Cirigliano:2017,Cirigliano:2018JHEP}.

The long-range type-II mechanism is related to the dimension-six and -seven LNV operators in low-energy EFT~\cite{Cirigliano:2017}. Thus, relations between the coefficients $\mathcal{C}^{\Rmnum{2}}$s  and Wilson coefficients are 
\begin{subequations}
\label{eq:append_C2}
    \begin{align}
    \mathcal{C}^{\Rmnum{2}}_{1}&=C^{(6)}_{VL}+C^{(6)}_{VR},\\
    \mathcal{C}^{\Rmnum{2}}_{2}&=C^{(6)}_{VL}-C^{(6)}_{VR},\\
    \mathcal{C}^{\Rmnum{2}}_{3}&=\frac{B}{m_e}(C^{(6)}_{SL}-C^{(6)}_{SR})+\frac{m^2_{\pi}}{\varv m_e}(C^{(7)}_{VL}-C^{(7)}_{VR}),
    \end{align}
\end{subequations}
where $B\simeq 2.8$ GeV denotes the quark condensate, related to the pion mass by $m_{\pi}^2=B(m_u+m_d)$, and $\varv=246$ GeV is the vacuum expectation value of Higgs field. It is noted that the LECs in $\mathcal{C}^{\Rmnum{2}}$s are absorbed into neutrino potentials~(see Eq.\eqref{eq:neutrino_potentials}), as usual. 

For the short-range type-III mechanism associated to the dimension-nine LNV operators in low-energy EFT~\cite{Cirigliano:2017}, the relations between $\mathcal{C}^{\Rmnum{3}}$s and Wilson coefficients are
\begin{subequations}
\begin{align}
    \mathcal{C}^{\Rmnum{3}}_{1L(R)}=&\frac{m_{\pi}^2}{\varv m_e}[g_2^{NN}(C^{(9)}_{2L(R)}+C^{(9)'}_{2L(R)})+g_3^{NN}(C^{(9)}_{3L(R)}+C^{(9)'}_{3L(R)})\nonumber\\
    &+g_4^{NN}C^{(9)}_{4L(R)}+g_5^{NN}C^{(9)}_{5L(R)}],\\
    \mathcal{C}^{\Rmnum{3}}_{1M}=&\frac{m_{\pi}^2}{\varv m_e}(g_6^{NN}C^{(9)}_V+g_7^{NN}\tilde{C}^{(9)}_V),\\
    \mathcal{C}^{\Rmnum{3}}_{2L(R)}=&\frac{m_{\pi}^2}{\varv m_e}\left(g_1^{\pi N}-\frac{5}{6}g_1^{\pi\pi}\right)\Bigg(C^{(9)}_{1L(R)}+C^{(9)'}_{1L(R)}\Bigg),\\
    \mathcal{C}^{\Rmnum{3}}_{2M}=&\frac{m_{\pi}^2}{\varv m_e}(g_V^{\pi N}C^{(9)}_V+\tilde{g}_V^{\pi N}\tilde{C}^{(9)}_V),\\
    \mathcal{C}^{\Rmnum{3}}_{3L(R)}
    =&\frac{1}{2\varv m_e}\Bigg[g_2^{\pi\pi}(C^{(9)}_{2L(R)}+C^{(9)'}_{2L(R)})+g_3^{\pi\pi}(C^{(9)}_{3L(R)}+C^{(9)'}_{3L(R)})\nonumber\\
    &-g_4^{\pi\pi}C^{(9)}_{4L(R)}-g_5^{\pi\pi}C^{(9)}_{5L(R)}\Bigg],
\end{align}
\end{subequations}
where $g_{i}^{NN(\pi\pi)}$ with $i\in\{2,3,4,5\}$, $g_{6(7)}^{NN}$ and $g_V^{\pi N}(\tilde{g}_V^{\pi N})$ are LECs, which can only be extracted from experimental data or lattice QCD calculations, but can be estimated by naive dimensional analysis. We have $g_{i}^{NN}=\mathcal{O}((4\pi)^2)$, $g_6^{NN}=g_7^{NN}=g_1^{\pi N}=g_1^{\pi\pi}=g_V^{\pi N}=\tilde{g}_V^{\pi N}=\mathcal{O}(1)$, $g^{\pi\pi}_{i}=\mathcal{O}(\Lambda_{\chi}^2)$  with $\Lambda_\chi \sim 1{~\rm GeV}$.  Using our NMEs in \nuc{Xe}{136}, together with the half-life limit from KamLAND-Zen and the corresponding phase-space factors, we present the lower limits on possible new physics scales, which are shown in Fig.~\ref{fig:ScaleLimits}.


%

\end{document}